**Rapid Communication**

# "The Impending Demise of Comet C/2013 US10 (Catalina)"

Ignacio Ferrín,
Institute of Physics,
Faculty of Exact and Natural Sciences,
University of Antioquia,
Medellin, Colombia, 05001000
ignacio.ferrin@udea.edu.co

Number of pages    12

Number of Figures    7

Number of Tables    2

**Abstract**

We have developed a 3-step criterion to decide if a comet coming from the Oort Cloud will disintegrate. If we apply this criterion to comet C/2013 US10 we find that the probability of disintegration is 92%. The Secular Light Curve of this comet exhibits complexity beyond current scientific understanding, suggesting that our knowledge of cometary science is incomplete.

**1. Introduction**

It is an observational fact that a fraction of comets coming from the Oort Cloud have a tendency to split or disintegrate. Stefanik (1966), Harwit (1968), Weissman (1980) and Sekanina (1984) investigated this tendency. Sekanina (2002) analyzed some of these cases in more detail. Since 1996 many of these comets have been studied observationally with CCDs and thus our knowledge and statistics have improved as can be seen from Table 1 and Figure 1, were 28 confirmed cases are listed. From 1996 to 2014 (19 years) there were 22 disintegrating comets, at a rate of 1.15 comets/year. This is a large and significant value. These are the comets that were in dead row.

On 2013 comet C/2012 S1 ISON, an Oort Cloud member, was declared "the comet of the century". In spite of this it disintegrated on November 28th of 2013 (Battams and Knight, 2013). Sekanina (2014) estudied it extensively. Using the same methodology used in this work, the comet was predicted to disintegrate up to six months in advance (Ferrín, 2014a, 2014c, papers submitted on 2013, June 20$^{th}$ and October 20th).

In the present work, we will study the Secular Light Curve (SLC) of comet C/2013 US10, another Oort Cloud member, and reach a similar conclusion: The comet is predicted to disintegrate. Additionally, the SLC exhibits complexity beyond current scientific understanding.

Mobberley (2015) presents a series of images of comet C/2013 US10 in his web site. The comet looks bright and robust and has a long tail  However, appearances can be misleading (Figure 0). Nothing can be deduced from these images that would hint to a dramatic final.

**2. Secular Light Curves of Comets, SLCs**

In a series of papers we have been developing the concept of Secular Light Curves of Comets (SLCs) (Ferrín, 2005; 2006; 2007; 2008; 2010a; 2010b; 2014a; 2014b; 2014c; Ferrín et al., 2012; 2013), a scientific way to show the brightness history of a comet. The SLCs can be represented in several renditions, for example $m(1,R,0)$ vs $\log R$, (where $m(\Delta,R,\alpha)$ means the magnitude at the comet-Earth distance $\Delta$, the Sun-comet distance, R, and the phase angle, $\alpha$). We could also plot $m(1,1,0)$ vs $t - T_q$, where $T_q$ is the time of perihelion, or $m(1,1,0)$ vs $\log R$. All these are equivalent rendition of the SLC.

In this work we adopt the *envelope of the dataset* as the correct interpretation of the observed brightness (Ferrín, 2005). Many physical effects affect comet observations. All these effects diminish the captured photons coming from the comet, and the observer makes an error downward, toward fainter magnitudes (one of the most common is the insufficient aperture of measurement error). There are no corresponding physical effects that could increase the perceived brightness of a comet. Thus, the *envelope* is the correct interpretation of the data. The envelope represents an ideal observer, with an ideal telescope and detector, with a large measuring aperture capable of extracting the whole flux from the comet, in an ideal atmosphere.

27 SLCs appear in the *Atlas of Secular Light Curves of Comets (*Ferrín, 2010). A full interpretation of the SLCs is given there and will not be repeated here. To carry out this investigation we reduced 1356 photometric observations of comets following that methodology.

**3. SLC Comet C/1973 E1 Kohoutek = 1973 f**
Before showing the SLC of comet C/2013 SU10, we will show that of comet Kohoutek for comparison. This well-observed comet, dubbed also "the comet of the century" (Anonymous, 1973), was expected to be very bright in 1973. However according to media and scientific reports, the comet "fizzled" and was a disappointment. Using the powerful tool provided by the SLC methodology, we can now investigate if these reports were based on scientific evidence. In Figure 2 we present the SLC of comet C/1973 E1 Kohoutek. We find robust activity with all the characteristics of a normal comet (confirmation, the 27 SLCs presented in the *Atlas*). Therefore, in reality comet Kohoutek did not "fizzle", since it never showed signs of fizzling. It was all a big misunderstanding by the media, the public and the scientist involved. Of course, at that time, 1973, we did not have the tool provided by the SLCs. The SLC methodology is capable of debunking this mistaken and often repeated conclusion.

The same plot is presented in another rendition in Figure 3, this time showing m(1,1) vs Log R. We see that after the SDE the comet continues brightening because it continues approaching the Sun. A bare nucleus with no activity would have had a flat light curve implying a constant absolute magnitude. This is not observed.

Figures 2 and 6 and those from the *Atlas*, show that the "normal" SLC of an Oort Cloud comet is basically composed of two straight lines separated by a Slope Discontinuity Event, SDE. After the SDE the normal Oort Cloud comet continues increasing in brightness up to perihelion. Disintegrating comets, on the other hand, halt their brightness after the SDE and show a standstill in magnitude (Figure 4). The standstill may last for months (confirmation, Ferrín, 2014a, 2014b, 2014c).

**4. Secular Light Curve of Comet C/2013 US10**
The SLC of this comet is shown in Figure 4 and it is like no other. Three databases were used. The MPCOBS database is a photometric-astrometry catalog of observations (Williams, 2015). Although the dispersion of the data is large, the envelope is clearly distinguishable and represents those astrometric observations with a sufficiently large aperture to extract the whole flux from the comet. We also used the COBS database administered by Jure Zakrajsek (2015). Lehmann (2015) published a light curve of comet

Catalina and his data is also plotted in Figure 4. An internet site that is useful to pre-view light curves in the form of m(Δ,R,α) vs time, is managed by Yoshida (2015).

Figure 4 shows that the comet exhibits a SDE and then a first standstills (SS) of 153 days duration. Then there is a second SDE and a second SS of 206 days duration. The total time expended in SS is 359 d, 6 d short of a complete year. This is a very long time, the longest of any of this group of comets. The two SSs are separated vertically by 1.0 magnitude. The first standstill starts far away from the Sun, at -8.07±0.04 AU and the second at -5.72±0.04 AU where water ice cannot yet sublimate.

## 5. Diameter of nucleus

It is possible to calculate a minimum diameter for the nucleus. Ferrín, (2010) has shown that there is a maximum amplitude of the secular light curve of a comet, $A_{SECULAR} = m_{NUCLEUS}(1,1,0) − m_{COMA}(1,1,0) \leq 11.6±0.2$ magnitudes. From Figure 4 we deduce that $m_{COMA}(1,1,0) = 4.3±0.2$, thus $m_{NUCLEUS}(1,1,0) \leq 15.9±0.3$. If the geometric albedo is taken as $p_V = 0.04$, then $D_{NUCLEUS} \geq 4.4±0.6$ km, a relatively large comet.

## 6. Criterion for Disintegration

In Table 1 we list several parameters of 28 comets that have disintegrated: the name of the comet, the perihelion distance, the eccentricity, and the status. The status comes from examining the SLC: SDE = Slope Discontinuity Event. SS = standstill. D = Disintegration. All 28 comets have a D consistent with the fact that this is a table of disintegrated comets. In 12 of these comets, a SS was observed, and in 10 of these, the SDE was also observed. Sekanina (1984) studied some of the early comets of this Table. Hergenrother (2010) discovered the disintegration of comet C/2009 R1 (McNaught). Gary Kronk, Toni Cook and Jacub Cerny (personal communication), discovered four members of the group.

In Figure 5 we examine the correlation between the eccentricity, e, and the perihelion distance, q. We see that the distribution of points is very sharp, with the standard deviation in the fourth decimal place, σ= 0.00045.

With this information, it is possible to stablish a criterion for disintegration: (1) The comet must exhibit a SDE. (2) The comet must exhibit a SS. In addition (3), the eccentricity must be confined to a very narrow range of values. The e-restriction is $0.99861 < e < 1.0008549$. All 12 comets for which we have information that have fulfilled these conditions have disintegrated (confirmation, Table 1). A few normal Oort Cloud comets lie in this region of the e vs q diagram, however they fail to fulfill the SED+SS requirement and have failed to disintegrate.

Figure 5 also shows the location of comet C/2013 US10 in the middle of the plot. Therefore, there is evidence to conclude that this comet is going to disintegrate since it fulfills the three conditions. If this comet failed to disintegrate, the probability of disintegration would have been p=1/13=7.7%. Thus the probability of disintegration is 100%-p=92.3%.

Another way to calculate the disintegration probability is the database of disintegration distances which contains only 14 data points (Ferrín, 2014c). If we set the perihelion distance of C/2013 US10 (R=0.82 AU) as the disintegration distance, then the probability of disintegration is 50%. That is, half of the observed comets had disintegrated by the time they reached this distance.

The fact that the range of eccentricities is so narrow and the fact that these comets disintegrated, suggests that they are new to the interior solar system, and thus that they must be dynamically new. Thus a new definition of dynamically new object might be $0.99861 < e(DN) < 1.0008549$.

Finally, Figure 6 shows several of these comets plotted together along the SLCs of normal Oort Cloud comets. Normal Oort Cloud comets have SLCs composed of two straight lines separated by the SDE, after which they continue brightening up to perihelion. Disintegrating comets follow a different path. Comet Catalina is bright and had to be moved down by 4 magnitudes to make a better comparison with the other disintegrated objects. Notice how closely Catalina is following the path of Elenin.

## 7. Observational conditions

Unfortunately, the observational conditions are not favorable. Table 2 gives the Date, RA, DEC, the distance to the Sun, R, and the elongation of the comet. In September 2015 the comet is approaching the Sun and entering the Sun's glare. At perihelion with R=0.82 AU, where the probability of disintegration is maximum, the comet is at the other side of the Sun. So it will be difficult to know if the comet has survived. However, at the end of November the comet is coming out of the Sun's glare and should be visible or invisible. Observers are encouraged to follow up on this comet and help determine its fate.

## 8. Conclusions

(1) Comets of the Oort Cloud announce that they are going to disintegrate by exhibiting the SDE + SS + e-restriction signatures. Of the 12 comets with these signatures in Table 1, 12 have disintegrated. The e-restriction is very sharp: $0.99861 < e < 1.0008549$ as concluded from the standard deviation, $\sigma = 0.00045$, thus these objects must all be dynamically new.

(2) Additionally the fact that the comet is following the path of disintegrating comets and not the path of normal Oort Cloud comets, implies that it is very probable that the comet will disintegrate. This path resembles that of comet C/2010 X1 Elenin, a comet that disintegrated.

(3) Based on the fact that 12 comets that fulfilled these three conditions all disintegrated, it is concluded that the probability of disintegration of C/2013 US10 is close to 92% although in another calculation we find 50%.

(4) We have been able to set a lower limit for the diameter of this comet. We find $D_{NUCLEUS} \geq 4.4$ km.

(5) This work raises many questions. For example: a) what is the physical meaning of the Slope Discontinuity Event? b) What is the meaning of the standstill? C) Why does this comet exhibit two standstills? d) Why is the eccentricity restricted to such a narrow range of values? The answers to these questions require of a complex model of a comet and are beyond the scope of this paper, which focuses only on the predictive power of the SDE+SS signature and path on the reduced magnitude vs Log R plot.

(6) A conclusion of this work is that the SLC of this comet exhibits complexity beyond current scientific understanding, suggesting that our knowledge of cometary science is incomplete.

**Acknowledgements**

We acknowledge with thanks the comet observations from the COBS Comet Observation Database contributed by observers worldwide and used in this research. The same acknowledgement is valid for the MPCOBS database with hundreds of individuals contributing their work to the database.

**REFERENCES**
Anonymous, 1973. Special Report. Kohoutek: Comet of the Century. Time magazine, December 17, p. 36.
Battams, K., Knight, M., 2013. Comet C/2012 S1 ISON. CBET 3731,
Ferrín, I., 2005. Secular Light Curve of Comet 28P/Neujmin 1, and of Comets Targets of Spacecraft, 1P/Halley, 9P/Tempel 1, 19P/Borrelly, 21P/Grigg-Skejellerup, 26P/Giacobinni-Zinner, 67P/Chruyumov-Gersimenko, 81P/Wild 2". Icarus 178, 493-516.
Ferrín, I., 2006. Secular Light Curve of Comets: 133P/Elst-Pizarro. Icarus, 185, 523-543.
Ferrín, I., 2007. Secular Light Curve of Comet 9P/Tempel 1. Icarus, 187, 326-331.
Ferrín, I., 2008. Secular Light Curve of Comet 2P/Encke, a comet active at aphelion. Icarus, 197, 169-182. http://arxiv.org/ftp/arxiv/papers/0806/0806.2161.pdf
Ferrín, I., 2010a. "Secular Light Curve of Comet 103P/Hatley 2, the next target of the Deep Impact EPOXI Mission". PSS, 58, 1868-1879. http://arxiv.org/ftp/arxiv/papers/1008/1008.4556.pdf
Ferrín, I., 2010b. "Atlas of Secular Light Curves of Comets". PSS, 58, 365-391. http://arxiv.org/ftp/arxiv/papers/0909/0909.3498.pdf
Ferrín, I., Hamanova, H., Hamanova, H., Hernandez, J., Sira, E., Sanchez, A., Zhao, H., Miles, R., 2012. The 2009 Apparition of Methuselah Comet 107P/Wilson-Harrington: A Case of Comet Rejuvenation? PSS, 70, 59-72. http://arxiv.org/ftp/arxiv/papers/1205/1205.6874.pdf
Ferrín, I., Zuluaga, J., Cuartas, P., 2013. "The location of Asteroidal Belt Comets on a Comets' evolutionary diagram: The Lazarus Comets". MNRAS, 434, 1821-1837. http://arxiv.org/ftp/arxiv/papers/1305/1305.2621.pdf
Ferrín, I., 2014a. "The Location of Oort Cloud Comets C/2011 L4 Panstarrs and C/2012 S1 (ISON), on a Comets' Evolutionary Diagram". MNRAS, 442, 1731-1754. http://arxiv.org/ftp/arxiv/papers/1306/1306.5010.pdf
Ferrín, I., 2014b. Three Predictions: Comet 67P/Churyumov-Gerasimenko, comet C/2012 K1 Panstarrs, and comet C/2013 V5 Oukaimeden, P&SS, 99, 171-175. http://arxiv.org/ftp/arxiv/papers/1403/1403.1781.pdf
Ferrín, I., 2014c. The impending demise of comet C/2012 S1 (ISON). P&SS, 96, 114-119. http://arxiv.org/ftp/arxiv/papers/1310/1310.0552.pdf
Ferrín, I., 2015. The Medellin Scale, a number-color code for comets. http://astronomia.udea.edu.co/cometspage/MEDELLINCOLORCODEB.xhtml
Hergenrother, C., 2010. Comet C/2009 R1 McNaught).
http://transientsky.wordpress.com/2010/06/27/new-comet-blues-comet-mcnaught-falls-short/

Harwit, M., 1967.  Astrophy. J., Spontaneously Split Comets.  151, 789-790.
Lehmann, T., 2015.  The light curve of comet C/2013 US10 Catalina.
    http://postimg.org/image/k2y7674j7/
Mobberley, M., 2015.  Images of C/2013 US10.
    http://martinmobberley.co.uk/images/2013us10_20150913_0910_mpm.jpg
Pittich, E. M., 1972.  Splitting and Sudden outbursts of comets as indicators of non-Gravitational effects.  In The Motion, Evolution of Orbits, and Origin of Comets; Proceedings from IAU Symposium no. 45, held in Leningrad, U.S.S.R., August 4-11, 1970. Edited by Gleb Aleksandrovich Chebotarev, E. I. Kazimirchak-Polonskaia, and B. G. Marsden. International Astronomical Union. Symposium no. 45, Dordrecht, Reidel, p.283
Sekanina, Z., 1984.  Disappearance and disintegration of comets.  Icarus, 58, 81-100.
Sekanina, Z., 2002.  What Happened to Comet C/2002 O4 (Hoenig)?  International Comet Quarterly, Vol. 24, pp. 223-236.
Sekanina, Z., 2014.  Disintegration of Comet C/2012 S1 (ISON) Shortly Before Perihelion: Evidence from Independent Data Sets.  http://arxiv.org/pdf/1404.5968v6.pdf
Weissman, P. R., 1980.  Physical Loss of Long Period Comets.  Astron. Astrophys., 85, 191-196
Williams, G., 2015. Administrator.  Minor Planet Center depository of astrometric observations, http://www.minorplanetcenter.net/db_search
Yoshida, S., 2015.  Home Page.  http://www.aerith.net/
Zakrajsek, J., 2015.  COBS Administrator.  http://www.cobs.si

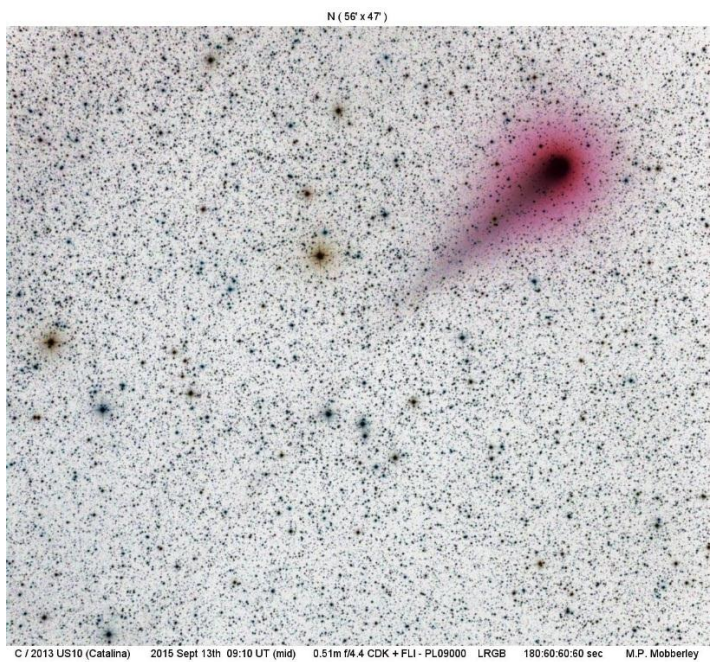

**Figure 0.** Mobberley (2015) presents a series of images of comet C/2013 US10 in his web site. This negative reproduction is one of them. Using the "Medellin Scale" (Ferrín, 2015) this comet would be classified as a "green comet". This is a healthy, strong, robust, active comet, whose color comes from a large production of CN (blue) and C2 (yellow), giving a final hue of green. However, appearances can be misleading. Nothing in this image hints to a dramatic final. (Reproduced with permission).

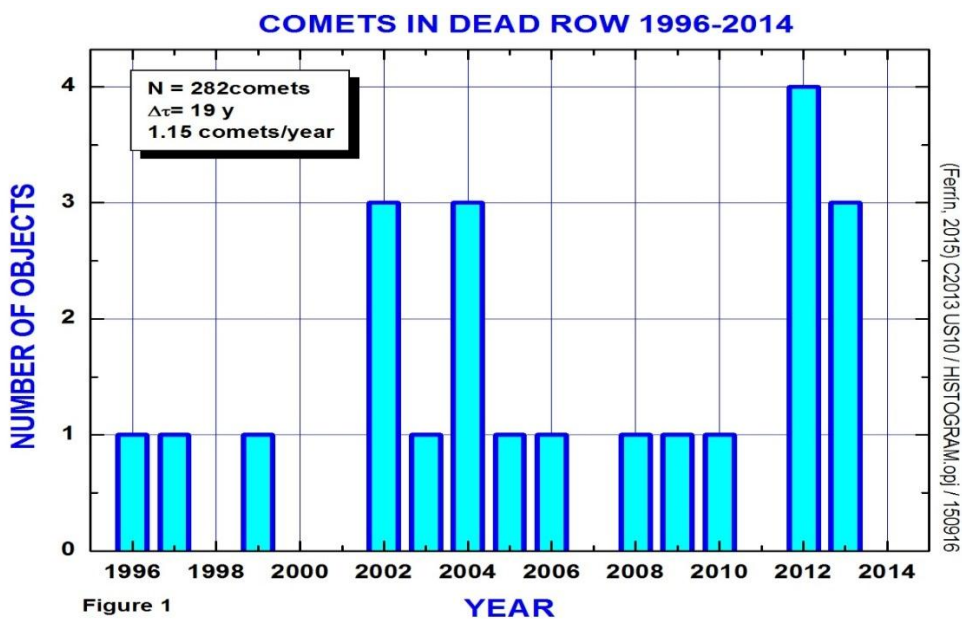

**Figure 1.** Histogram of number of Oort Cloud disintegrating comets vs year. From 1996 to 2014 (19 years) there were 22 disintegrating comets, a rate of 1.15 comets/year.

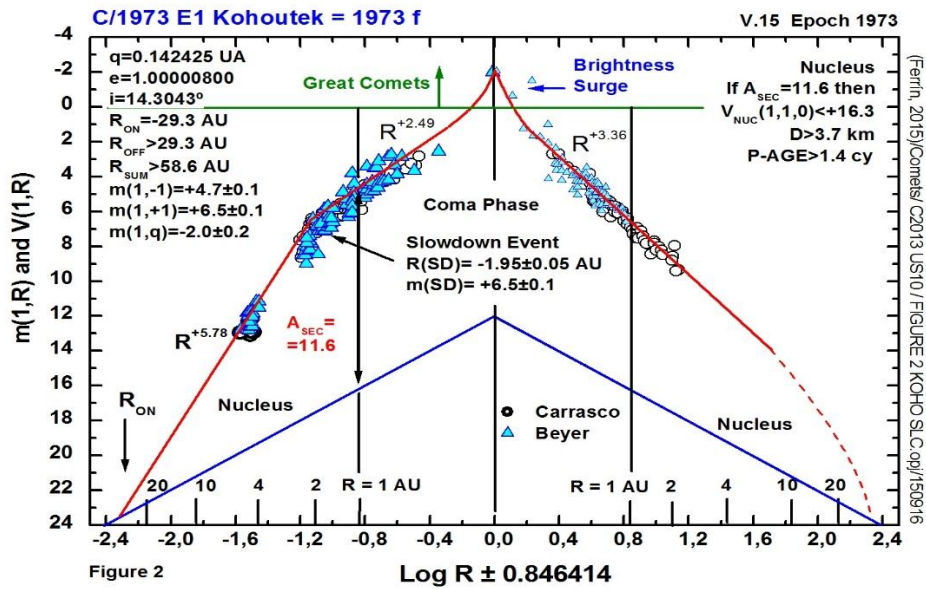

**Figure 2**. The SLC of comet C/1973 E1 Kohoutek. The vertical axis plots m(1,R) and the horizontal axis Log R. Negative logs are positive. The (-) sign indicates only observations pre-perihelion. The pyramid line at the bottom of the plot is the $R^{-2}$ law of a bare inactive nucleus. The SLC is entirely normal, and exhibits a SDE and a continuous increase in brightness after the SDE. Notice the power law $R^{+5.78}$ before SDE and $R^{+2.49}$ after the SDE. Thus, this comet did not "fizzled" as concluded by media and scientific reports of the epoch. The SLC methodology is capable of debunking this mistaken and widespread conclusion.

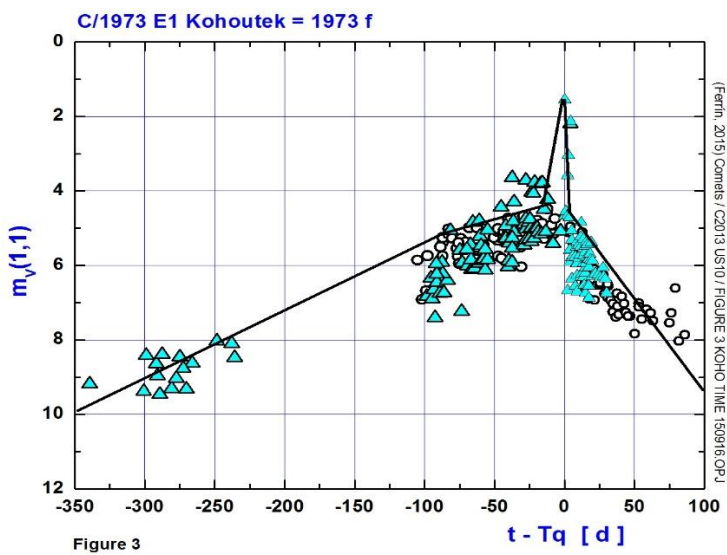

**Figure 3.** The SLC of comet C/1973 E1 is shown in another rendition. The vertical axis plots the absolute magnitude m(1,1). The horizontal axis plots the time with respect to perihelion. The result is the same. The brightness continues increasing after the SDE. *If there were no activity*, the absolute magnitude would be the same along the orbit, and this plot would be a flat horizontal line (the $R^{-2}$ law of a bare inactive nucleus).

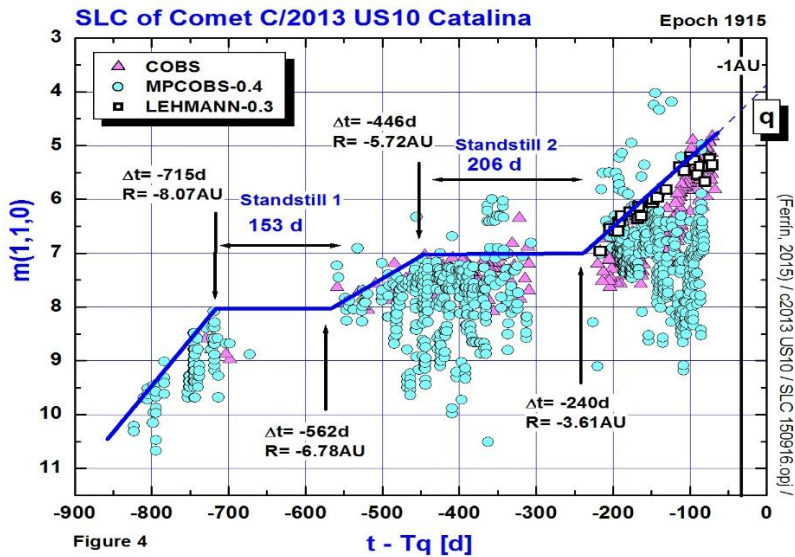

**Figure 4.** The very odd SLC of comet C/2013 US10 (Catalina). The vertical axis is the absolute magnitude. The horizontal axis the time with respect to perihelion. It exhibits two SDEs followed by two SSs. This behavior is entirely different to the behavior of a normal Oort Cloud comet, shown in Figures 2, 3, 6 and the *Atlas*, composed of two straight lines of different slope separated by a SDE.

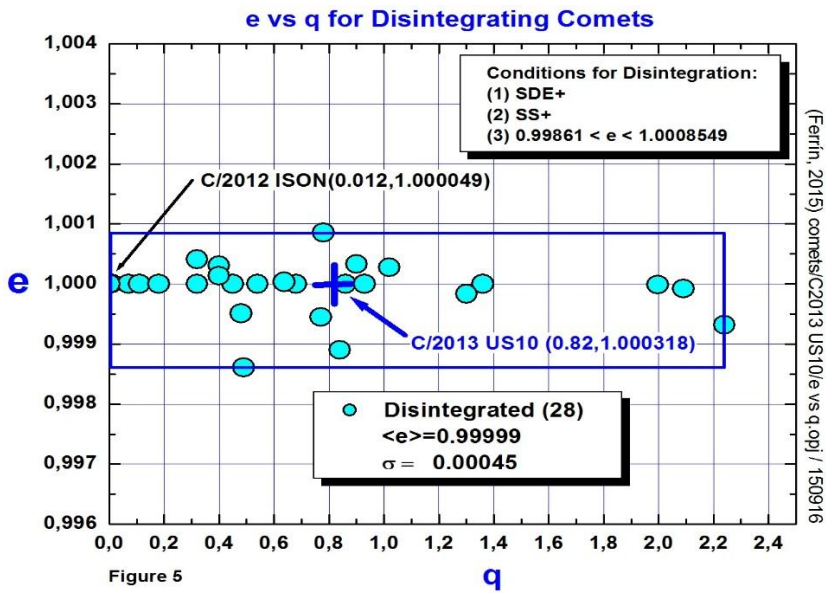

**Figure 5.** e vs q diagram, where e is the eccentricity and q the perihelion distance. Three conditions seem to predict disintegration: 1) a SDE. 2) A standstill. 3) An eccentricity limited by $0.99861 < e < 1.0008549$. In this Figure, we see that the e-restriction is very sharp, as can be derived from the standard deviation significant in the fourth decimal place. Since these comets disintegrated, this must be their initial visit to the inner solar system suggesting that they are dynamically new. Thus a new definition of dynamically new comet might be $0.99861 < e(DN) < 1.0008549$.

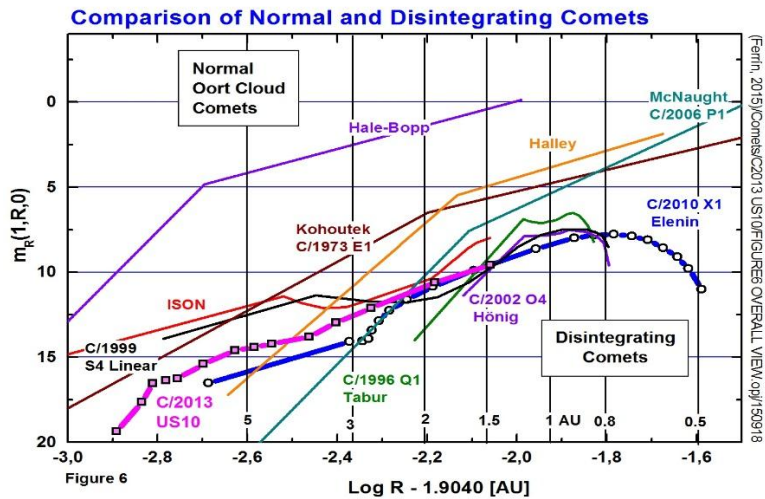

**Figure 6.** A comparison is made of normal Oort Cloud comets with disintegrating comets. The plot displays the envelopes of these comets. Normal Oort Cloud comets show two linear laws separated by a SDE. After the SDE they continue increasing in brightness up to perihelion. Disintegrating comets follow a different path. Shortly after the SDE, they disintegrate. Comet C/2013 US10 (Catalina) was moved down by 4 magnitudes to better compare it with other disintegrated comets. This comet is not following the path of normal Oort Cloud comets, and thus it is probable that it will disintegrate. The comet is following very closely the path of C/2010 X1 Elenin, a comet that disintegrated.

```
Table 1. Comets that have disintegrated.
------------------------------------------------------------------
#  Comet                           q[AU]  e           status
------------------------------------------------------------------
01 C/1887 B1 Headless comet        0.005  1,0000000   ---+--+D
02 C/1897 U1 Perrine= 1897 III     1.36   1,0000000   ---+--+D
03 C/1925 X1 Ensor = 1926 III      0.32   1,0000000   ---+--+D
04 C/1953 X1 Pajdusakova 1954 II   0.07   1,0000000   ---+--+D
05 C/1957 U1 Latyshev-Wild-Burnham 0.54   1,0000000   ---+--+D
06 C/1974 V2 Bennet 1974XV         0.86   1,0000000   ---+--+D
07 C/1996 Q1 Tabur                 0.84   0,9989006   ---+--+D
08 C/1997 N1 Tabur                 0.40   1,0001344   ---+--+D
09 C/1999 S4 LINEAR                0.77   0,9994490   SDE+SS+D
10 C/2002 O4 Hönig                 0.78   1,0008549   ---+--+D
11 C/2002 O6 SWAN                  0.49   0,9986079   ---+--+D
12 C/2002 O7 LINEAR                0.90   1,0003300   ---+SS+D
13 C/2003 H1 LINEAR                2.24   0,9993207   SDE+SS+D
14 C/2004 S1 Van Ness              0.68   1,0000000   ---+--+D
15 C/2004 R2 ASAS                  0.11   1,0000000   SDE+SS+D
16 C/2004 V13 SWAN                 0.18   1,0000000   ---+--+D
17 C/2005 K2 LINEAR                0.54   1,0000000   ---+--+D
|8 C/2006 VZ13 LINEAR              1.02   1,0002752   SDE+SS+D
19 C/2008 J4 McNaught (headless)   0.45   1,0000000   ---+--+D
20 C/2009 R1 McNaught              0.40   1,0003074   SDE+SS+D
21 C/2010 X1 Elenin                0.48   0,9995109   SDE+SS+D
22 C/2012 V1 PANSTARRS             2.09   0,9999224   ---+--+D
23 C/2012 S1 ISON                  0.012  1,0000049   SDE+SS+D
24 C/2012 T5 Bressi                0.32   1,0004086   SDE+SS+D
25 C/2012 CH17 MOSS                1.30   0,9998326   ---+SS+D
26 C/2013 G5 CATALINA              0.93   1,0000000   ---+--+D
27 C/2013 A1 Siding Spring         1.40   1,0004167   SDE+SS+D
28 C/2013 V5 Oukaimeden            0.63   1,0003320   SDE+SS+D
------------------------------------------------------------------

TABLE 2. Ephemeris of Comet C/2013 US10 (MPC)
YYYY MM DD  RA           DEC          R[AU]   ELON   MAG
2015 09 16  14 51 34.1  -47 49 19     1.365   66.7   6.5
2015 09 26  14 41 07.6  -41 23 50     1.236   53.0   6.3
2015 10 06  14 35 14.7  -36 12 10     1.113   40.3   6.0
2015 10 16  14 31 18.2  -31 46 01     1.002   28.2   5.6
2015 10 26  14 28 02.8  -27 41 46     0.911   16.9   5.3
2015 11 05  14 24 59.5  -23 40 08     0.848    8.8   5.0
2015 11 15  14 22 12.1  -19 25 19     0.823   13.4   4.8 perihelion
2015 11 25  14 19 58.6  -14 41 26     0.841   24.6   4.7
2015 12 05  14 18 32.3  -09 05 20     0.899   37.1   4.8
2015 12 15  14 17 38.8  -01 53 44     0.988   50.6   4.8
2015 12 25  14 16 18.7  +08 15 59     1.096   65.6   4.9
2016 01 04  14 12 12.9  +23 43 27     1.217   83.1   4.9
2016 01 14  13 58 11.1  +46 28 49     1.346  102.3   5.0
2016 01 24  12 55 04.1  +71 47 57     1.478  115.2   5.5
```